\documentclass{aastex}
\usepackage[onecolumn]{emulateapj5}
\newcommand{\dpdp}[2]{\frac{\partial #1}{\partial #2}}
\newcommand{\bm}[1]{\mbox{\boldmath $#1$}}

\begin{document}
\title{Generation of Seed Magnetic Field around First Stars:\\Effects of
Radiation Force}
\author{Masashi Ando\altaffilmark{1}, Kentaro Doi\altaffilmark{2}, and Hajime Susa\altaffilmark{3}}
\affil{Department of Physics, Konan University, Okamoto, Kobe, Japan}
\altaffiltext{1}{a03cb004z@gmail.com}
\altaffiltext{2}{mn921009@center.konan-u.ac.jp}
\altaffiltext{3}{susa@konan-u.ac.jp}
\begin{abstract}
We investigate seed magnetic field generation in the early universe by
radiation force of first stars. 
In a previous study with the steady assumption, large amplitudes($\sim
10^{-15}$G for first stars, $\sim 10^{-11}$G for QSOs) are predicted.  
In this study, we formulate this issue in an unsteady framework.
Then, we consider a specific model of magnetic field
generation around a very massive first star. Consequently, we (1) find that the steady
assumption is not valid in realistic situations, and (2) obtain much
smaller magnetic field strength than predicted by Langer et al. In
addition, we find that the momentum transfer process during photoionization is more
important than Thomson scattering. The resultant magnetic flux density
around the first star is $\la 10^{-19}$G. This seed magnetic field will not
affect subsequent star formation in the neighborhood of first stars.
\end{abstract}
\keywords{early universe---HI\hspace{-.1em}I regions ---magnetic fields ---radiative transfer}

\section{Introduction}
One of the most important current issues in cosmology is understanding the star formation history of the universe, especially the
early universe. Owing to progress in the last decade, mainly due to
theoretical studies, we now expect the very first stars in the early universe to be very massive ($\ga 100M_\odot$), compared to those in
present-day galaxies\citep[e.g.,][]{Bromm02,NU01,Abel02,
Yoshida06a}.

The chief reason that the first stars are very massive is the lack of
heavy elements in the primordial gas, since the heavy elements are efficient
coolants. Theoretical studies suggest that when
the metallicity of the star forming gas exceeds $10^{-5} {\rm to} 10^{-6}Z_\odot$,
the star formation mode transits from very massive star formation to normal
star formation\citep{schneider03,omukai05}. Thus, metallicity is the key factor for the transition of
the star formation mode.

Another possible parameter is magnetic field strength. In the
present-day star-forming molecular cloud, energy density of the magnetic
field is comparable to gravitational or turbulent energy density, while
it is assumed that the magnetic field is weak in the primordial gas.
Magnetic field in present-day galaxies basically suppresses the star
formation process.

The effects of magnetic field on the early star formation episode in the
universe have been studied by several authors. 
The coupling between magnetic field and pure primordial gas has been investigated\citep{maki04,maki07}. They found that
the magnetic field is always frozen in the gas in so far as the magnetic field
strength is not strong enough to prevent gravitational
contraction. 
Based on the frozen-in assumption, it is found that
bipolar outflows emerge in cases $B > 10^{-9}$G is satisfied at $n_{\rm
H}=10^3 {\rm cm^{-3}}$ \citep{machida06}.
\citet{tan_blackman} suggest that magnetorotational Instability(MRI) is
activated in the accretion disk surrounding proto first stars in cases where $B
> 10^{-10}$G at $n_{\rm H}=10^3 {\rm cm^{-3}}$, which results in
efficient angular momentum transfer in the disk. It is also reported by
\citet{schleicher} that gravitational contraction of the primordial gas
is delayed in cases where the comoving field strength is larger than $10^{-10}$G.

Therefore, it is
crucial to determine the magnetic field strength that exists in the
environment of the star-forming gas cloud in the early universe in order to
understand the transition of the star formation mode.
Magnetic field strength at the star formation site in the early universe
is basically provided by the field generated prior to nonlinear
structure formation \citep[e.g.,][]{harrison1,MR72,turner,ichiki}, as
well as the field generated by the Biermann battery effect\citep{biermann}
 during the structure formation \citep[e.g.,][]{xu} or reionization
 \citep{gnedin}.
The predicted field strength in these models is within the  range
of $10^{-16}{\rm G} {\rm to} 10^{-20}{\rm G}$, which is too weak to affect star formation. Another possibility is that the magnetic field spewed out
from the formed stars \citep[e.g.,][]{BRS73}. In cases where the field
strength is as large as $\sim$1 G like normal stars, a simple flux freezing
assumption again leads to $\sim 10^{-16}$G at intergalactic medium(IGM) densities.

On the other hand, \citet{langer} suggest a mechanism to generate a large
amplitude of seed magnetic field as large as $10^{-11}$G, which could
affect the star formation in the early universe.
They consider HI\hspace{-.1em}I regions formed by luminous sources
like QSOs or first stars. The radiation force field due to the photons emitted
by these objects is anisotropic because of the clumpiness of
surrounding media. This anisotropy causes charge separation in the
HI\hspace{-.1em}I region, which generates a large scale eddy current
as well as the magnetic fields.
However, the assumption in this theory is idealized.
In their theory, anisotropy of radiation is dealt with as perturbation, and
the HI\hspace{-.1em}I region is assumed to be steady.
In reality, dense shadows are generated behind the nearby dense halos. In
addition, HI\hspace{-.1em}I regions formed by QSOs or first stars are
actually unsteady, and ionization fronts never reach
the Str$\ddot{\rm o}$mgren radius \citep{shapiro,kitayama,yoshida07}.

In this paper, we investigate this magnetic field generation process by
radiation force in nonlinear and unsteady framework along the lines of theory of
\citet{langer}. We consider a specific model of magnetic field
generation around a very massive first star and assess whether the generated magnetic
field affects the subsequent star formation. 

In Section 2, we describe the basic equations. 
A brief description and the setup of our models are described in Section 3. 
In Section 4, we show the results of our calculation, and Section 5 is
devoted to discussion.
Finally, in Section 6, we conclude our research.

\section{Basic equations}
\subsection{Magnetic field generation}
\label{mag_generation}
In this section, we describe the basic equations of magnetic field
generation. To begin with, the force balance on a single electron is
written as follows:
\begin{eqnarray}
0&=&-e\bm{E}
    +\frac{e}{\sigma_{\rm c}}\bm{j}
    +\bm{f}_{\rm rad}
    -\frac{\bm{\nabla}p_{\rm e}}{n_{\rm e}}
    -\frac{e}{c}\bm{v}_{\rm e}\times\bm{B}\label{eq:eom_electron}
\end{eqnarray}
Here, $e,~\bm{E},~\bm{B},$ and $~\bm{j}$ denote the elementary charge, electric
field strength, magnetic flux density, and electric current density,
respectively. $p_{\rm e}, n_{\rm e},$ and $\bm{v}_{\rm e}$ represent the
pressure, number density and velocity of electrons as
fluid, respectively. $\sigma_{\rm c}$ is the conductivity and $\bm{f}_{\rm rad}$ is
the radiation force on an electron.

The fluid velocity $\bm{v}$ is defined as  
\begin{equation}
\bm{v}\equiv\frac{n_{\rm p}m_{\rm p}\bm{v}_{\rm p}+n_{\rm e}m_{\rm
 e}\bm{v}_{\rm e}}{n_{\rm p}m_{\rm p}+n_{\rm e}m_{\rm e}}\simeq
 \bm{v}_{\rm p}
\end{equation}
where $n_{\rm p}$,$m_{\rm p}, m_{\rm e}$, and $\bm{v}_{\rm p}$ denote the
number density of protons, the mass of
protons, the mass of electrons, and the velocity of protons, respectively.
Thus, the velocity of electrons is approximated as $\bm{v}_{\rm e}\simeq
\bm{v} -\bm{j/}(en_{\rm e})$. Substituting this notation into
Equation(\ref{eq:eom_electron}), we obtain
\begin{eqnarray}
0&=&-e\bm{E}
    +\frac{e}{\sigma_{\rm c}}\bm{j}
    +\bm{f}_{\rm rad}
    -\frac{\bm{\nabla}p_{\rm e}}{n_{\rm e}}
    -\frac{e}{c}\bm{v}\times\bm{B}+\frac{1}{cn_{\rm e}}\bm{j}\times\bm{B}\label{eq:eom_electron2}
\end{eqnarray}

Since we consider the expanding HII region around a first star, we have
to solve the following photoionization rate equation for electrons.
\begin{eqnarray}
\frac{\partial n_{e}}{\partial t}  + \bm{\nabla} \cdot(n_{e} \bm{v}_{e})
 &=& \Gamma - \alpha n_{e} n_{p}\label{eq:cont_e}
\end{eqnarray}
Combining Equation (\ref{eq:cont_e}) with the similar equation for
protons, we obtain the charge conservation equation
\begin{equation}
\dpdp{\rho}{t}+\bm{\nabla}\cdot\bm{j}=0,\label{eq:charge_cons}
\end{equation}
where charge density $\rho$ and current density $\bm{j}$ are related to
the number density of electrons and protons as $\rho\equiv e\left(n_{\rm p}-n_{\rm e}\right)$ and $\bm{j}\equiv e\left(n_{\rm p}\bm{v}_{\rm p}-n_{\rm e}\bm{v}_{\rm e}\right)$.
We also need the set of Maxwell equations: 
\begin{eqnarray}
\bm{\nabla}\cdot\bm{E}=4\pi\rho,~~~~~\bm{\nabla}\times\bm{E}=-\frac{1}{c}\dpdp{\bm{B}}{t},~~~~~
\bm{\nabla}\cdot\bm{B}=0,~~~~~\bm{\nabla}\times\bm{B}=\frac{4\pi}{c}\bm{j}+\frac{1}{c}\dpdp{\bm{E}}{t}\label{eq:Maxwell_eq}
\end{eqnarray}
Equations
(\ref{eq:eom_electron2}),(\ref{eq:cont_e}),(\ref{eq:charge_cons}) 
and (\ref{eq:Maxwell_eq}) combined with hydrodynamic equations are the fundamental
equations for this problem.

In order to obtain the equation that describes the growth of magnetic
flux density, we apply $-(4\pi\sigma_{\rm c}/ec)\bm{\nabla}\times$ to
both sides of Equation (\ref{eq:eom_electron2}). We have
\begin{eqnarray}
0=-\frac{4\pi}{c}\bm{\nabla}\times\bm{j} +
 \frac{4\pi\sigma_{\rm c}}{c}\bm{\nabla}\times\bm{E}+
\frac{4\pi\sigma_{\rm c}}{c^2}\bm{\nabla}\times\left(\bm{v}\times\bm{B}\right)
-\frac{4\pi\sigma_{\rm c}}{c^2}\bm{\nabla}\times\left(\frac{\bm{j}\times\bm{B}}{n_{\rm e}}\right)-\frac{4\pi\sigma_{\rm c}}{cen_{\rm e}^2}\bm{\nabla}n_{\rm e}\times\bm{\nabla}p_{\rm e}
-\frac{4\pi\sigma_{\rm c}}{ce}\bm{\nabla}\times\bm{f}_{\rm rad}\nonumber
\end{eqnarray}

Using the second and fourth equations of Maxwell Equations
(\ref{eq:Maxwell_eq}), we have 
\begin{eqnarray}
\frac{1}{c^2}\dpdp{^2\bm{B}}{t^2} -\bm{\nabla}^2\bm{B} +
 \frac{4\pi\sigma_{\rm c}}{c^2}\dpdp{\bm{B}}{t}=
\frac{4\pi\sigma_{\rm c}}{c^2}\bm{\nabla}\times\left(\bm{v}\times\bm{B}\right)
-\frac{4\pi\sigma_{\rm c}}{c^2}\bm{\nabla}\times\left(\frac{\bm{j}\times\bm{B}}{e
					 n_{\rm e}}\right)
-\frac{4\pi\sigma_{\rm c}}{cen_{\rm e}^2}\bm{\nabla}n_{\rm e}\times\bm{\nabla}p_{\rm e}
-\frac{4\pi\sigma_{\rm c}}{ce}\bm{\nabla}\times\bm{f}_{\rm rad}
\end{eqnarray}
In this expression, $4\pi\sigma_{\rm c}$ denotes
the inverse of the timescale that describes the convergence of electron
velocity to the terminal velocity. This timescale is quite short, since
$\sigma_{\rm c}\simeq 6.5\times 10^6 T^{3/2}$e.s.u.\citep{lang}. It is as short as a picosecond in cases where we consider photoionized
 gas. 
Thus, the first and second terms on the left-hand side of the above
 equation are negligible compared to the third term, since the typical
 timescale is the lifetime of the source star which is as long as
 Myr, and the typical length scale is kpc. Finally, we obtain 

\begin{eqnarray}
\dpdp{\bm{B}}{t}=\bm{\nabla}\times\left(\bm{v}\times\bm{B}\right)
-\frac{c}{e n_{\rm e}^2}\bm{\nabla}n_{\rm e}\times\bm{\nabla}p_{\rm e}
-\frac{c}{e}\bm{\nabla}\times\bm{f}_{\rm rad}\label{eq:B_growth}
\end{eqnarray}

Here we also omitted the Hall current term which is proportional to
$\bm{j}\times \bm{B}$, since this is a higher order term than the others, in
cases where we consider the generation of very weak magnetic field. The first
term on the right-hand side denotes the advection term. If the second and
third terms are neglected, the frozen-in condition is satisfied. The second
term describes the well-known Biermann battery term\citep{biermann}, 
and the third
term represents the radiation term. Thus, the non-zero rotation of the radiation
force field always introduces the source term of magnetic field generation.
It is worth noting that the gravitational force cannot be the source of
magnetic field generation, since it is described by the gradient of a
scalar potential.

We also mention that the final Equation (\ref{eq:B_growth}) does not
contain the terms directly originated from the electric current. In
fact, Equation (\ref{eq:B_growth}) is independent of
conductivity, $\sigma_{\rm c}$. The time derivative of magnetic field simply
comes from Faraday's induction equation, i.e., the second equation of
Maxwell Equation (\ref{eq:Maxwell_eq}). Thus, the magnetic field
generation by the radiation force in this paper 
can be regarded as the consequence of electromagnetic induction effects.
The electric field with non-zero rotation 
is generated so as to cancel the radiation force. 
Thus, it is very natural that the second equation of
Maxwell Equation (\ref{eq:Maxwell_eq}) coincides with Equation
(\ref{eq:B_growth}), if we assume $e\bm{E}=\bm{f}_{\rm rad}$ and neglect
the advection term and Biermann term in Equation (\ref{eq:B_growth}).
We also note that a similar problem was considered in detail for the
problem of field generation due to radiation forces in accretion disks
around black holes \citep{BB77,Contopoulos,BL02,Contopoulos2,Christodoulou}.

Aside from the magnetic field generation described by Equation
(\ref{eq:B_growth}), we can assess the charge separation which
drives the electric current. Taking the divergence of Equation
(\ref{eq:eom_electron2}) combined with the charge conservation Equation
(\ref{eq:charge_cons}) and the first Maxwell Equation(\ref{eq:Maxwell_eq}),
we have
\begin{eqnarray}
\dpdp{\rho}{t}+4\pi\sigma_{\rm c}\rho=\sigma_{\rm c}\bm{\nabla}\cdot\left(
-\frac{1}{c}\bm{v}\times\bm{B}+\frac{1}{e n_{\rm e}
c}\bm{j}\times\bm{B}-\frac{\bm{\nabla}p_{\rm e}}{en_{\rm e}}+\frac{1}{e}\bm{f}_{\rm rad}\right)
\end{eqnarray}
The second term on the left hand-side is much larger than the first
term, since $1/(4\pi\sigma_{\rm c}) \ll t$ is always
satisfied. Therefore, the charge density is assessed as 
\begin{eqnarray}
\rho=\frac{1}{4\pi}\bm{\nabla}\cdot\left(-\frac{1}{c}\bm{v}\times\bm{B}+\frac{1}{e n_{\rm e}c}\bm{j}\times\bm{B}-\frac{\bm{\nabla}p_{\rm e}}{en_{\rm e}}+\frac{1}{e}\bm{f}_{\rm rad}\right)\label{eq:charge_density}
\end{eqnarray}
Thus, separated charge is piled up where the divergence of the force
field is large. For instance, if we consider only
$\bm{f}_{\rm rad}$ on the right-hand side of Equation (\ref{eq:charge_density}), 
$\bm{\nabla}\cdot\bm{f}_{\rm rad}$ becomes negative in the neighborhood of the
ionization front, since $\bm{f}_{\rm rad}$ changes very rapidly across the
ionization front. Thus, electrons will accumulate around the ionization front.

\subsection{Momentum transfer from radiation to electrons}
In order to follow the generation of magnetic field, we need to evaluate
the radiation force on electrons, $\bm{f}_{\rm rad}$. 
Two elementary processes could be potentially
important for this momentum transfer. First of all, Thomson scattering could
contribute to $\bm{f}_{\rm rad}$. 
We use the formal solution of the radiation transfer equation for
$\bm{f}_{\rm rad,T}$ as follows:
\begin{eqnarray}
\bm{f}_{\rm rad,T}  = \frac{\sigma_{\rm T}}{c}\int_0^{\nu_{L}}\bm{F}_{0\nu}d\nu + \frac{\sigma_{\rm T}}{c}\int_{\nu_{L}}^{\infty}\bm{F}_{0\nu}\exp\left[- \tau_{\nu_{L}}a\left(\nu\right)\right]  d\nu,\label{eq:momtr_T}
\end{eqnarray}
where $\bm{f}_{\rm rad,T}$ denotes the radiation force per single
electron due to Thomson scattering, 
$\sigma_{\rm T}$ is the cross section of Thomson scattering, 
$\bm{F}_{0\nu}$ is the unabsorbed energy flux density,
$\nu_{L}$ denotes the Lyman-limit frequency,
$\tau_{\nu_{L}}$ is the optical depth at Lyman limit regarding the
photoionization, and $a(\nu)$ denotes the frequency dependence of
photoionization cross section, which is normalized at Lyman limit,
i.e., $a(\nu_L)=1$ is satisfied.

Second, the photoionization process itself also transfers the photon momentum
to electrons. Its contribution to $\bm{f}_{\rm rad}$ is given as
\begin{eqnarray}
\bm{f}_{\rm rad,I}  = \frac{1}{2}\frac{n_{\rm HI}}{c n_{\rm e}}
\int_{\nu_{L}}^{\infty} \sigma_{\nu_{L} } a\left(\nu\right)
\bm{F}_{0\nu}\exp\left[- \tau_{\nu_{L}} a\left(\nu\right)\right]  d\nu, \label{eq:momtr_ion}
\end{eqnarray}
where $\sigma_{\nu_L}$ denotes the photoionization cross section at the
Lyman limit. We remark that the factor $1/2$ in the right-hand side of
Equation (\ref{eq:momtr_ion}) is due to the fact that the photon momentum
is equally delivered to protons and electrons. 
In fact, when electrons and protons are emitted by photoionization, its
momentum is basically delivered to surrounding electrons and protons by
Coulomb interaction. Since the number density and charge of each
particle is the same for electrons and protons, electrons gain half of the
momentum transferred from photons.
Because of the large inertia of protons,
we can also safely assume that only
electrons are accelerated by this momentum transfer process.

The photoionization rate $\Gamma$ in Equation (\ref{eq:cont_e}) is also
obtained by assuming the formal solution of the radiation transfer equation:
\begin{eqnarray}
\Gamma = n_{\rm HI}\int_{\nu_{L}}^{\infty}\sigma_{\nu_{L}}a(\nu)\frac{\bm{F}_{0\nu}}{h\nu}\exp\left[- \tau_{\nu_{L}}a\left(\nu\right)\right]  d\nu
\end{eqnarray}
In practice, the photon conserving scheme \citep{Abel99,Susa06} is used
when we solve Equation (\ref{eq:cont_e}), so that we can trace the
propagation of the ionization front properly.
\begin{figure}[t]
\begin{center}
\includegraphics[angle=0,width=7cm,bb=0 0 587 552]{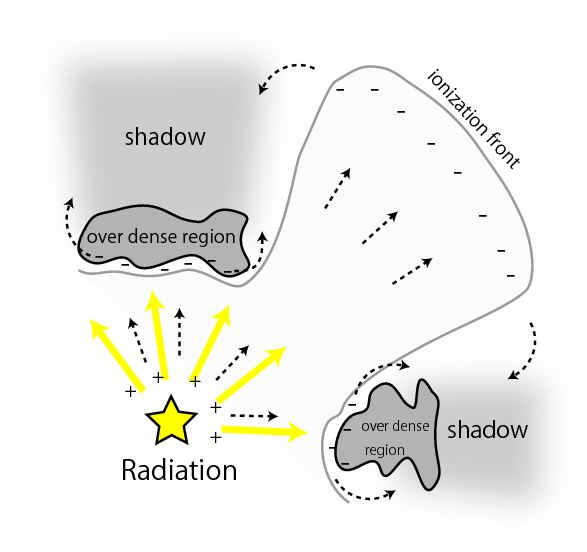}
\end{center}
\caption{Schematic view of the magnetic field generation
 process. Dashed arrows represent the direction of electron flow.}\label{fig1}
\end{figure}

\section{The Model}
\subsection{Description of the model}
We consider gas surrounding a first star. 
This ambient gas is ionized by the radiation
from the source, forming an HII region.
The momentum transfer from ionizing photons to
electrons leads to a slight charge separation in the HII region.
It creates a weak electric field balancing the radiation force.

The electric field and the distribution of charge cannot be isotropic,
because the propagation speed of the ionization front is anisotropic due
to the clumpiness of the surrounding media. 
As illustrated in Figure\ref{fig1}, overdense regions in
the neighborhood of the radiation source form shadows behind themselves that
result in anisotropic radiation flux  as well as the electric
field/charge distribution. Consequently, a transverse electric field along
the ionization front is generated. This field could drive eddy currents
surrounding the boundary between the shadowed and unshadowed regions. Thus,
the magnetic field is expected to be generated at this boundary region.
\subsection{Setup}
We consider a first star of $500M_{\odot}$($L =
5\times 10^{40}$ ${\rm erg} {\rm s}^{-1}$,~$t_{\rm age} = 2 \times 10^{6}$yr, $T_{\rm eff}=10^5$K \citep{schaerer}), at redshift $z \simeq 20$. 
Since the ionization front breaks out the small host halos with $M_{\rm
halo}\lesssim 10^7M_\odot$\citep{kitayama}, we consider the expansion of the
HII region in intergalactic space.
The initial number density of the ambient gas in intergalactic space is
$n_{0} = 10^{-2} {\rm cm^{-3}}$ at such redshift. It is also assumed
that the gas is neutral when the source star is turned on.

\begin{figure}
\begin{center}
\includegraphics[width=12cm,bb=0 0 1053 528]{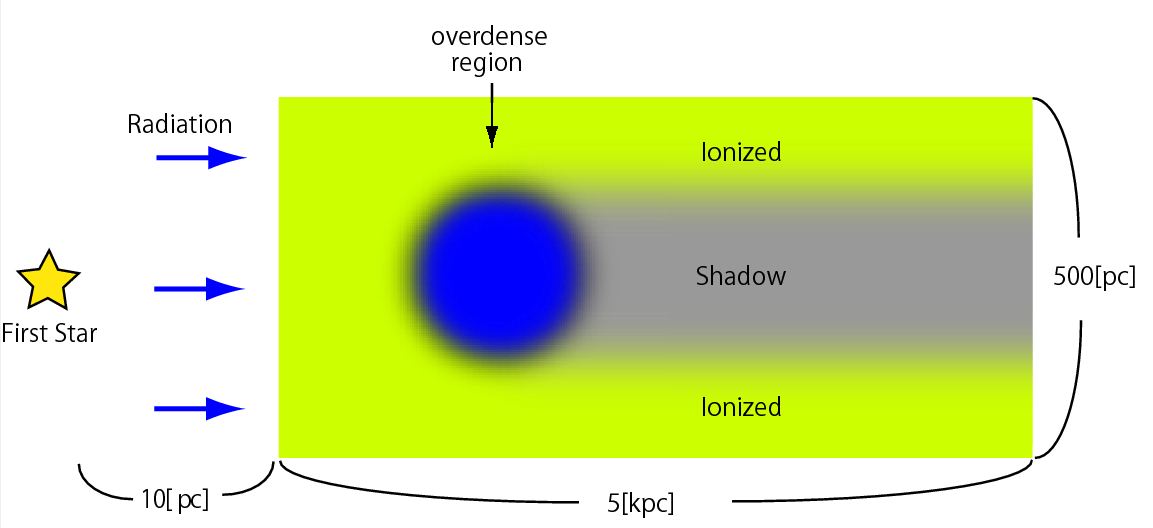}
\end{center}
\caption{Schematic view of the computational domain. The shadowed
 region is formed behind the overdense region , while the other region
 is exposed to the radiation field.}\label{fig2}
\end{figure}

As shown in Figure\ref{fig2}, we follow the growth of magnetic flux
density in a two-dimensional model.
The computational domain is a strip of 500pc$\times$5kpc.
The radiation flux flows from the left along the long edge of the
domain. The assumed flux on the left edge is the flux 10pc away from
the source first star.
We set an overdense region of $n_0{\rm cm^{-3}}$, $D$kpc away from the
left edge of the simulated region, which casts a shadow behind it
(Figure\ref{fig2}). 
We tested four models 
~~A:$n_0=1{\rm cm^{-3}}/D=2$kpc,~~B:$n_0=1{\rm cm^{-3}}/D=200{\rm pc}$,~~
C:$n_0=10{\rm cm^{-3}}/D=200{\rm pc}$, and D:$n_0=10{\rm cm^{-3}}/D=2{\rm kpc}$.
The overdense region has a core-halo structure. The
diameter of the core ($n=n_0{\rm cm^{-3}}$) is 50pc, which is surrounded
by the envelope of $n\propto r^{-2}$, where $r$ is the radial distance
from the center of the overdense region.

In this paper, we assume that the gas is isothermal ($T=10^4$K),
which results in eliminating the Biermann battery effect. In addition, we
also assume that the gas is at rest, i.e. $\bm{v}=0$. This assumption is
valid when the ionization front is propagating in the intergalactic
space; however, it is not appropriate in cases where the dense gas clump is
photoevaporated because of the thermal pressure of the photoheated
gas. These effects are left for future work. In this paper, we focus on
the effects of radiation.

We also remark that the timescale for the ionization front to reach the
Str$\ddot{\rm o}$mgren radius
is  $1/ (\alpha n_{e})\simeq 1.2 \times 10^{7} $yr\citep{shu}, 
which is sufficiently longer than the lifetime of the source
star. Therefore, the HII region does not settle down to a steady state
during the lifetime of the source star\citep{yoshida07}. 

\section{Results}
\subsection{Orders of magnitude}
\label{OOM}
Before we move to the numerical results, it is useful to show the
expected magnetic field strength by rough calculations.
Following the isothermal and static fluid assumptions, Equation
(\ref{eq:B_growth}) is reduced to a very simple equation:
\begin{eqnarray}
\dpdp{\bm{B}}{t}=
-\frac{c}{e}\bm{\nabla}\times\bm{f}_{\rm rad}\label{eq:B_growth2}
\end{eqnarray}
This expression shows that the magnetic field is effectively generated
where the radiation force field has a large shear. It is obvious that $\bm{f}_{\rm
rad}$ should have a large derivative at the boundary between the
shadowed and unshadowed regions. Thus, we focus our attention on such boundary region.

The radiation force $\bm{f}_{\rm rad}$ is assessed by Equations
(\ref{eq:momtr_T}) and (\ref{eq:momtr_ion}) as follows:
\begin{eqnarray}
\bm{f}_{\rm rad} = \bm{f}_{\rm rad, T} + \bm{f}_{\rm rad, I}
 \sim \frac{\bm{F}_{0}}{c}
 \left( \sigma_{\rm T} + \frac{n_{\rm HI}}{2 n_{\rm e}} \sigma_{\nu_L}\right),
\end{eqnarray}
where $\bm{F}$ is the total energy flux coming from the source star.  
Thus, $B$ is evaluated as
\begin{equation}
B\sim \frac{1}{e}\cdot\frac{1}{\Delta r} \cdot \frac{L_{*}}{4\pi R^2}
 \cdot \frac{1}{2}\sigma_{\nu_L} t_{\rm age} 
\sim  10^{-18} {\rm G} 
\left(\frac{\Delta r}{10{\rm pc}}\right)^{-1}
\left(\frac{L_{*}}{5} \times 10^{40}{\rm erg} {\rm s}^{-1}\right)
\left(\frac{R}{2{\rm kpc}}\right)^{-2}
\left(\frac{t_{\rm age}}{2\times 10^6{\rm yr}}\right).\label{eq:B_order}
\end{equation}

where $\Delta r$ denotes the length scale that represents the sharpness
of the boundary between the shadowed and unshadowed regions.
Here we assume $n_{\rm HI}/n_{\rm e}\sim 1$ and the flux is not absorbed
significantly ($F\sim L_*/(4\pi R^2)$). This assumption is only valid in
the neighborhood of the ionization front. In the neutral region, dumping by
the absorption of ionizing radiation reduces the radiation
force. Therefore, Equation (\ref{eq:B_order}) should be regarded as an
upper limit of the generated magnetic field strength.

It is also worth noting that Thomson scattering
is negligible in neutral regions since the cross section is smaller than
photoionization by 7 orders of magnitude, although it could be
dominant in the highly ionized region. That is why the contribution from
$\bm{f}_{\rm rad, T}$ is omitted in Equation (\ref{eq:B_order}).

\begin{figure}
\vspace{-7cm}
\begin{center}
\includegraphics[width=24cm,bb=0 0 1030 1063]{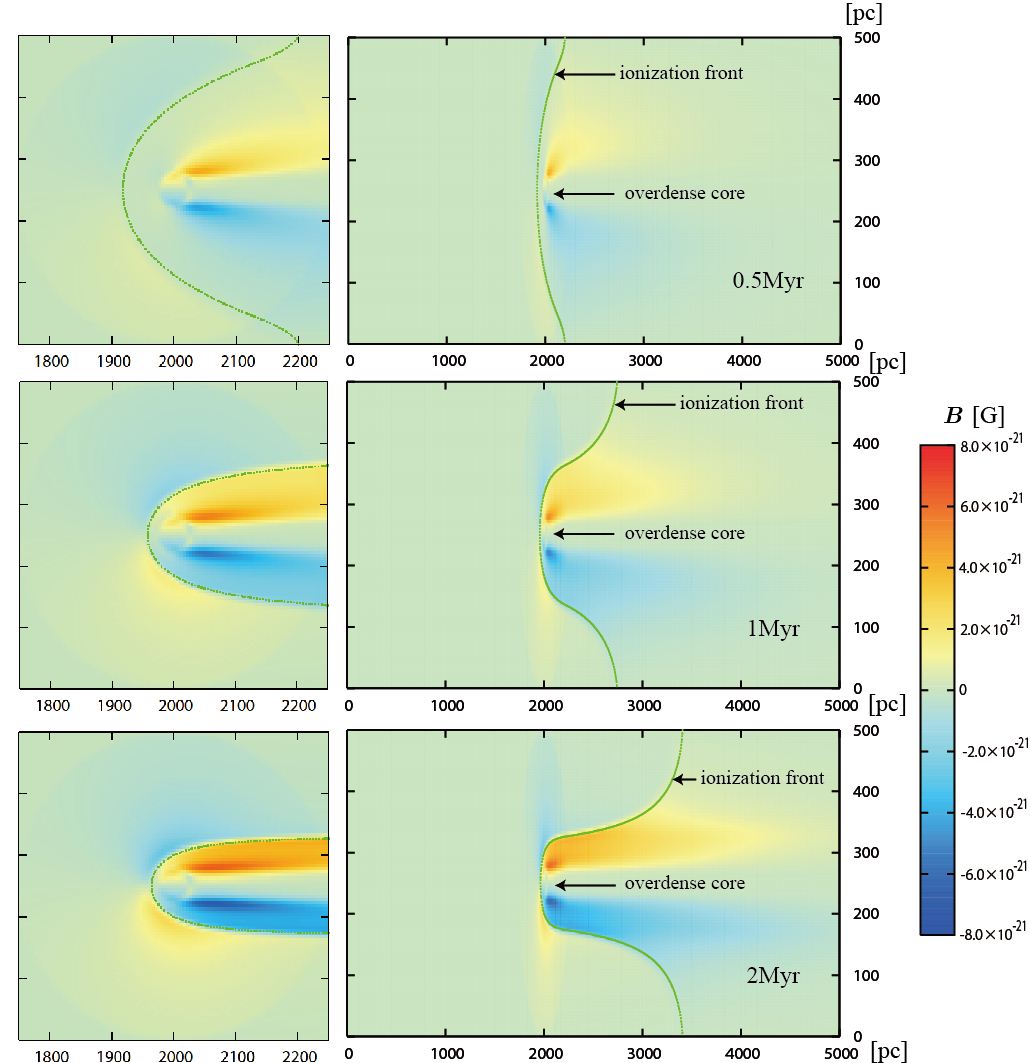}
\caption{
Color contour maps of the magnetic field strength at three snapshots at
 $0.5$, 1 and 2Myr. 
 The three snapshots on the left are extended views around the dense core.
 Green lines denote the position of the ionization front. Orange
 represents the region in which $\bm{B}$ is directed to the front side
 of the page, whereas blue is used for the opposite direction.}
\label{fig3}
\end{center}
\end{figure}
\begin{figure}
\begin{center}
\includegraphics[width=18cm,bb=0 0 1097 603]{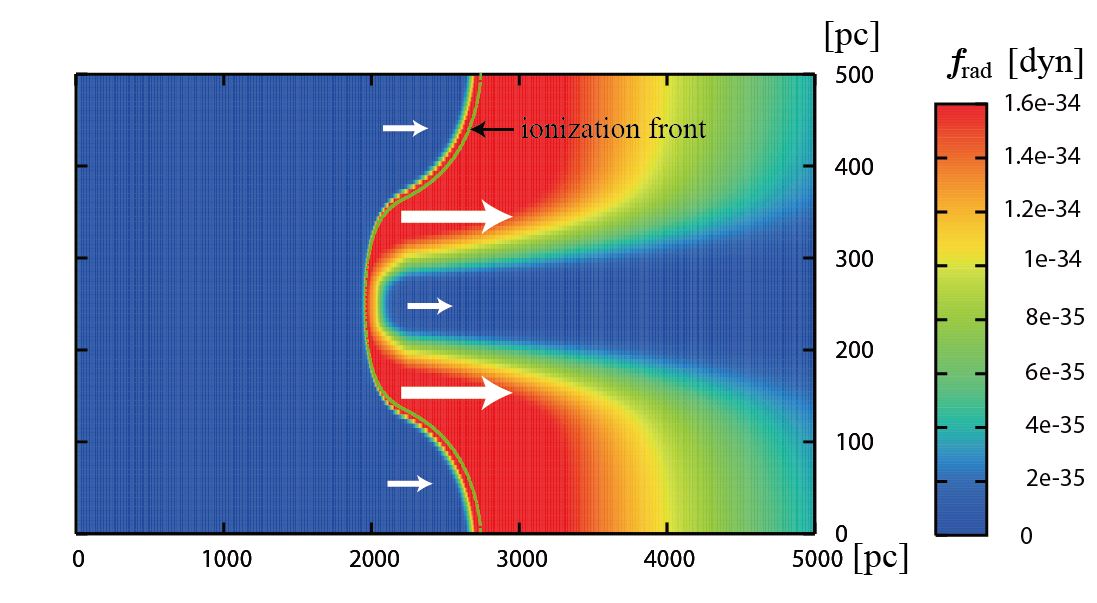}
\caption{Color contour map of the radiation force $f_{\rm rad}$ 1Myr after the ignition of the source star. White arrows
 schematically show the flow of electrons.}\label{fig:frad}
\end{center}
\end{figure}
\begin{figure}
\begin{flushleft}
\includegraphics[angle=0,width=14cm,bb=0 0 1144 795]{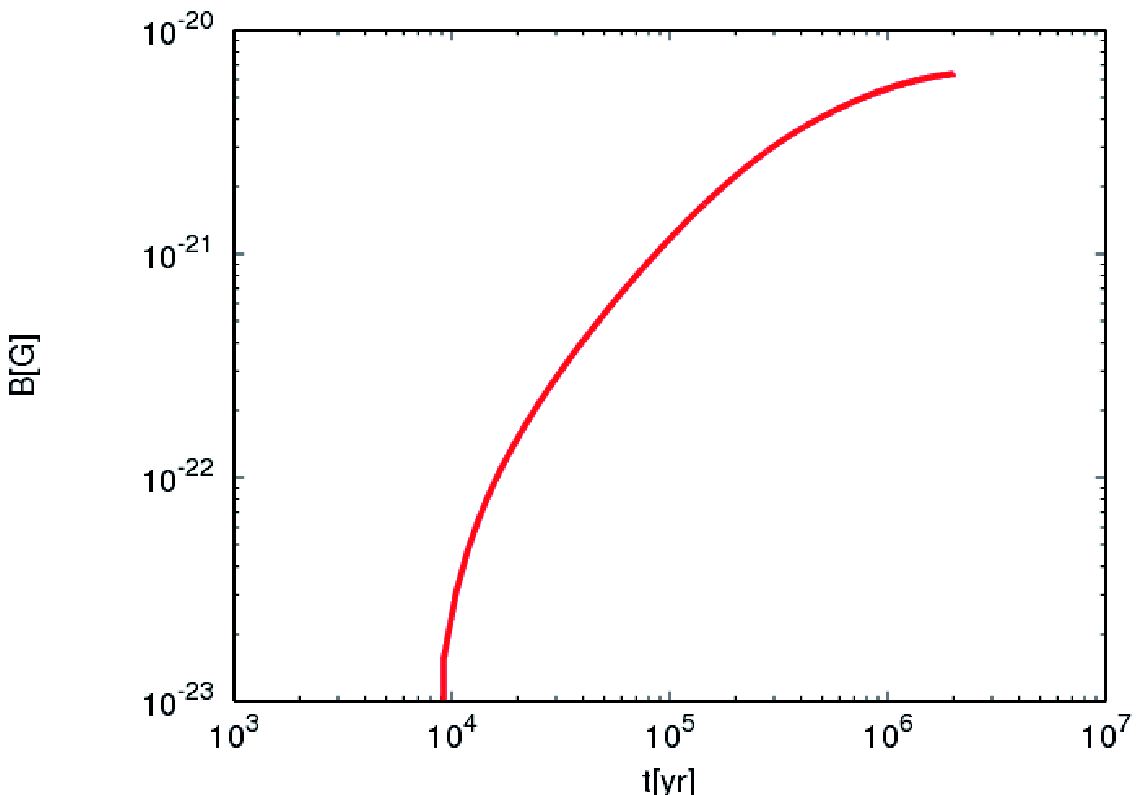}
\caption{Time evolution of the peak magnetic field strength is plotted for
 model A:$n_{\rm c} =1{\rm cm^{-3}}$, $D=2$kpc.}\label{fig:time_evol}
\end{flushleft}
\end{figure}
\begin{figure}
\vspace{-7cm}
\begin{center}
\includegraphics[width=24cm,bb=0 0 1025 1031]{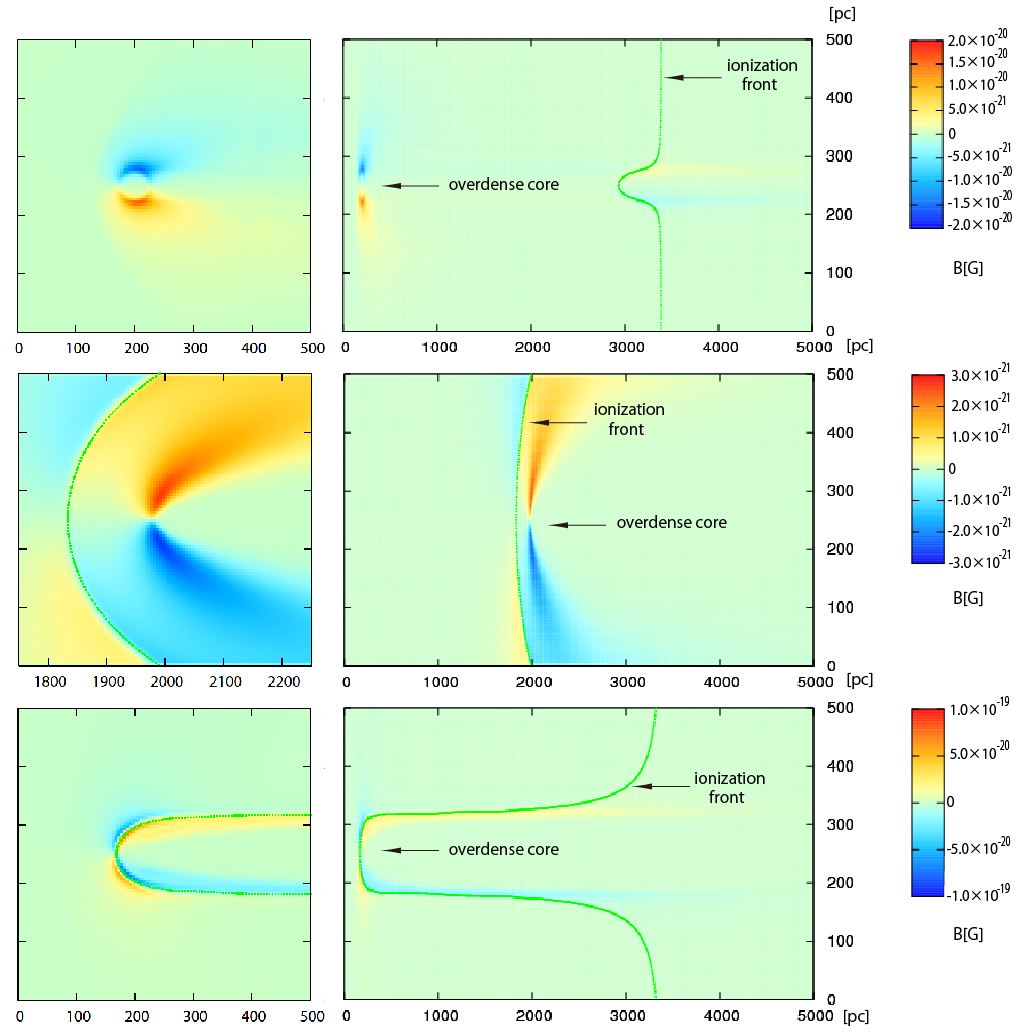}
\caption{Color contour maps of the magnetic field strength at 2Myr for
 model B:$n_{\rm c} =1{\rm cm^{-3}}$, $D=200$pc; 
model C:$n_{\rm c} =10{\rm cm^{-3}}$, $D=2$kpc; and model D:$n_{\rm c}
 =10{\rm cm^{-3}}$, $D=200$pc. The panels on the left represent
 extended views around the dense core. 
Green lines denote the position of the ionization front. Orange represents the region in which $\bm{B}$ is directed to the front side
 of the page, whereas blue is used for the opposite direction.}\label{fig:fig6}
\end{center}
\end{figure}

\subsection{Numerical results}
We integrate Equation (\ref{eq:B_growth2}) through the lifetime of the
source star ($2\times 10^{6} $yr) after it turned on.
Three snapshots for the distribution of magnetic field strength are
shown in Figure\ref{fig3} for model A:$n_{\rm c} =1{\rm cm^{-3}}$, $D=2$kpc. 
Three panels correspond to 0.5, 1 and 2Myr after the ignition of the source star, from top to bottom , respectively.
The radiation propagates from left to right in the panels, as the
ionization front(green curve). The ionization front is trapped at
the overdense core located 2kpc away from the left edge. Note that
the right panels are compressed along the direction of the light rays, since the
shadow is a quite elongated structure. The left panels have the same scale in both the
horizontal and vertical directions.

The orange area denotes the region where the magnetic field is
perpendicular to the plane of the page and directed to its front side.
In the blue region, the magnetic field points in the opposite direction. 
First of all, it is clear that the magnetic field strength peaks at the boundary
of the shadow cast by the dense core.  As discussed in Section \ref{OOM}, $\bm{f}_{\rm
rad}$ should vary significantly across the boundary between the
shadowed and unshadowed regions, since the radiation flux is very weak in the
shadow. 
In fact, as shown in Figure\ref{fig:frad}, where the color contour of
$\bm{f}_{\rm rad}$ is shown, $\bm{f}_{\rm rad}$ changes significantly behind the
dense core. Since the ionization front is trapped at the dense core,
the large shear of $\bm{f}_{\rm rad}$ is kept relatively long, while it is only
transient in the low-density region far from the dense core. 
Thus, the magnetic flux density reaches its peak just behind the dense 
core, since it has enough time to grow.

We also remark that the border of the shadow 
does not coincide with the ionization front. As shown in
Figure\ref{fig:frad}, $\bm{f}_{\rm rad}$ is larger behind the ionization front,
i.e., in the neutral region. This behavior indicates that the momentum
transfer from photons to electrons through the photoionization process
is more important than Thomson scattering 
since $\bm{f}_{\rm rad,I}$ is larger in
the neutral region, while Thomson scattering is important in the ionized region.

The time evolution of the peak magnetic field strength is plotted in Figure\ref{fig:time_evol}. The field strength immediately reaches the level of $\sim
10^{-21} $G after the ionization front hits the overdense region at
$t=2\times 10^5$yr. After that, the field strength grows gradually until
the source star dies. 

We also show three snapshots at 2Myr for three other models in Figure\ref{fig:fig6}.
These panels correspond to models B, C and D from top to bottom, respectively.

In model B the dense core is
located 10 times closer to the left edge than in the case of model A. We
expect 100 times stronger magnetic field strength following the
scaling relation of Equation (\ref{eq:B_order}). However, we have 
the maximum magnetic field strength of $\sim 2.0 \times 10^{-20}$G.
In this case, the ionization front is not trapped by the dense core. 
Consequently, the magnetic field does not have enough time to grow since the
shadowed region behind the dense core is feathered after the ionization
front passes through. Thus, the field strength does not follow the simple scaling relation.

Model C corresponds to the case where the density of the core is 10 times
larger than that in model A, while the distance $D$ is the same.
The peak magnetic field strength in this model is as large as $3.0 \times
10^{-21}$G, which is smaller than the case of model A by a factor of few.
Since the ionization front does not pass through even at the envelope of the
overdense region, the rotation of $\bm{f}_{\rm rad}$(i.e., the shear of
$\bm{f}_{\rm rad}$) around the dense core becomes relatively smaller
than that in model A. Therefore, the magnetic field strength is smaller
than the case of model A.

Finally, we describe the results of model D, which are shown in the bottom panels of Figure\ref{fig:fig6}.
In model D, the core is located 10 times closer, and is 10 times denser
than that in model A. As shown in Figure\ref{fig:fig6}, the ionization front
is trapped in front of the dense core because of the high enough core
density. It is also clear that the ionizing radiation flux is high enough
to ionize the envelope, unlike model C.
In this case, the maximal magnetic field strength is $\sim 10^{-19}$G. 

The parameters employed in models A to D  scan the reasonable range.
In fact, the actual size of the HII region of a 500$M_\odot$ first star
is as large as several kpc, and the size of the first halos is $\la 100$pc. The
density of the virialized halo is $\sim 1{\rm cm^{-3}}$ at
$z=20$. Thus, we can conclude that the magnetic flux density generated by radiation from
first stars is $\la 10^{-19}$G.

\section{Discussion}
We found that the generated magnetic field strength is $\sim 10^{-19}$G
at maximum if we adopt reasonable ranges for the parameters in our
model. This value is much smaller than those predicted by steady
theory\citep{langer}, although we take into account the additional
contribution by $\bm{f}_{\rm rad,I}$. If we consider only Thomson
scattering as done in \citet{langer}, we obtain a magnetic field of
$\sim 10^{-25}$G. This discrepancy comes from the steady assumption. 
In their study, the magnetic field strength was estimated using steady
equations, however;  as shown in Section \ref{mag_generation}, a much longer time than the age of the universe is needed to settle down to steady state in this situation.
Considering a realistic timescale, such as the age of the source star
$\sim 10^6$yr, we conclude that the steady assumption is wrong.

In the present work, the system is assumed to be isothermal ($=10^4$K) for
simplicity, which results in eliminating the Biermann battery term. 
In a more realistic situation, we have to consider the effects of radiation
hydrodynamics as well as the heating and cooling of gas by which we can compute the magnetic field generated by the battery effect\citep{biermann}.
In fact, the order of magnitude of magnetic field generated by the battery
effect is
\begin{equation}
B \sim \frac{c}{n_{e}^{2} e} \left(\frac{n_{e}}{\Delta r} \right)
 \left(\frac{p_{e}}{\Delta r} \right)\sin\theta \; t_{\rm age} \sim 2.0 \times
 10^{-18} \left( \frac{t}{2.0 \times 10^{6} {\rm yr}} \right) \left(\frac{\sin\theta}{0.1}\right)
 \left(\frac{\Delta r}{10 {\rm pc}} \right)^{-2} \left(\frac{T}{10^{4}}
						 \right) {\rm G}, \nonumber
\end{equation}
where $\Delta r$ denotes the typical length of $n_{\rm e}$ and $p_{\rm
e}$ change significantly and $\theta$ denotes the angle between
$\bm{\nabla}n_{\rm e}$ and $\bm{\nabla}T$. This expected field strength
is even larger than the flux density due to radiation pressure, although we have significant uncertainties in the above estimate.  
Thus, we need to include the Biermann battery term to assess the correct
magnetic field strength utilizing radiation hydrodynamic calculations,
which is left for future work.

We also remark that the magnetic field strength obtained in this
paper seems insufficient to affect star formation in the early
universe, since $B\sim 10^{-19}$G at IGM cannot account for
$10^{-10} {\rm to} 10^{-9}$G at $n=10^3{\rm cm^{-3}}$ required for jet formation \citep{machida06} and
MRI activation\citep{tan_blackman}. However, it is also worth noting
that a very recent study by \citet{schleicher10} indicates that turbulence in
the host halos of first stars \citep[e.g.,][]{Wise08} can amplify the magnetic field to
equipartition strength very quickly. If this is true, this might
suppress the fragmentation of gas disks formed at the center of
the star-forming gas \citep{machida08}, and help the MRI activation.

\section{Conclusion}
We investigate the magnetic field generation process by
radiation force in a nonlinear and unsteady framework along the line of the theory of
\citet{langer}. We consider a specific model of magnetic field
generation around a very massive first star. Using analytical and
numerical calculations, we find that the steady
assumption is not valid in realistic situations and obtain a much
smaller magnetic field strength than predicted by the steady theory. In
addition, we find that the momentum transfer process during photoionization is more
important than Thomson scattering. The resultant magnetic flux density
around first stars is $\sim 10^{-19}$G. This seed magnetic field does
not seem to affect subsequent star formation in the neighborhood of first stars.

\bigskip
We thank the anonymous referee for critical comments which helped to improve
the manuscript. We also thank N.Tominaga for fruitful discussions. This
work was supported in part by the Inamori Foundation, as well as Ministry of
Education, Science, Sports and Culture, Grant-in-Aid for Scientific
Research (C), 22540295. 


\end{document}